\documentclass[conference]{IEEEtran}

\usepackage[utf8]{inputenc} 
\usepackage{amsmath, amssymb, amsfonts} 
\usepackage{graphicx} 
\usepackage{tikz} 
\usepackage{cite} 
\usepackage{hyperref} 
\usepackage{xcolor} 
\usepackage{algorithm}
\usepackage{algorithmic}
\usepackage{microtype}
\usepackage{pgfplots}
\usepackage{physics}
\usepackage{subcaption}
\usepackage{listings}
\usepackage{xcolor}
\usepackage{tcolorbox}
\definecolor{darkgray}{rgb}{0.15,0.15,0.15}  
\usepackage{csquotes}

\definecolor{codegray}{rgb}{0.5,0.5,0.5}
\lstdefinestyle{mystyle}{
    language=Python,
    basicstyle=\ttfamily\footnotesize,
    keywordstyle=\color{blue}\bfseries,
    commentstyle=\color{gray},
    stringstyle=\color{red},
    numbers=left,
    numbersep=5pt,
    numberstyle=\tiny\color{codegray},
    frame=lines,
    breaklines=true
}

\pgfplotsset{compat=1.18}

\usetikzlibrary{arrows, automata, positioning, shapes.geometric, graphs, backgrounds, external}
\begin{document}

\title{QAOA in Quantum Datacenters: Parallelization, Simulation, and Orchestration }
\author{
    Amana Liaqat\textsuperscript{1}, Ahmed Darwish\textsuperscript{2}, Adrian Roman\textsuperscript{1}, and $^*$Stephen DiAdamo\textsuperscript{2} \\[1ex]
    \textsuperscript{1}Qoro Quantum, London, England \\ 
    \textsuperscript{2}Qoro Quantum, Munich, Germany \\     
    $^*$stephen@qoroquantum.de    
}

\maketitle

\maketitle

\begin{abstract}

Scaling quantum computing requires networked systems, leveraging HPC for distributed simulation now and quantum networks in the future. Quantum datacenters will be the primary access point for users, but current approaches demand extensive manual decisions and hardware expertise. Tasks like algorithm partitioning, job batching, and resource allocation divert focus from quantum program development. We present a massively parallelized, automated QAOA workflow that integrates problem decomposition, batch job generation, and high-performance simulation. Our framework automates simulator selection, optimizes execution across distributed, heterogeneous resources, and provides a cloud-based infrastructure, enhancing usability and accelerating quantum program development. We find that QAOA partitioning does not significantly degrade optimization performance and often outperforms classical solvers. We introduce our software components---Divi, Maestro, and our cloud platform---demonstrating ease of use and superior performance over existing methods.

\end{abstract}


\section{Introduction}
To scale quantum computing infrastructure to practical levels, both in the near and long term, networked systems are essential. In the near term, quantum computers can be integrated with classical High Performance Computing (HPC) infrastructure, forming quantum datacenters~\cite{alexeev2024quantum, mohseni2024build}. These datacenters offer users unified access to classical and quantum resources through a single interface. In the long term, quantum systems are expected to evolve into fully networked clusters, where quantum computers share entanglement over quantum networks to enable distributed quantum computing.

In the current near-term scenario, quantum computers in datacenters are often not connected via quantum networks. Instead, they function as distributed systems in the traditional sense, relying on classical connections. While this setup offers performance gains~\cite{carrera2024combining, mineh2023accelerating, chen2024noise}, it also imposes limitations. Without quantum entanglement between devices, qubit resources cannot be effectively shared, limiting scalability. Nevertheless, even classical networking can provide runtime improvements, though the specific benefits depend heavily on the application and how computational tasks are managed~\cite{moretti2024enhanced, giortamis2024orchestrating}.

As quantum algorithms grow in complexity, the need for more robust automation tools becomes evident. To date, most proof-of-concept implementations have relied on manual processes, requiring users to handle hardware allocation, simulator selection, scheduling, parallelization, and batch job management. This places a heavy burden on developers, detracting from their ability to focus on algorithm design. To fully leverage available infrastructure and scale applications efficiently, it is critical to offload these tasks to automated systems. Without automation, the increasing complexity of quantum workflows will make manual management infeasible, slowing progress in the field.

Our proposed solution, aligned with efforts like~\cite{schulz2022accelerating, mohseni2024build}, integrates high-performance simulation and hardware execution under a unified interface. This approach allows users to focus on solving problems without needing to manage infrastructure complexities. While some solutions already exist, there remains a gap in the ecosystem when it comes to optimizing specific applications for massively parallel and distributed systems. Developers still face fundamental questions: How should quantum algorithms be parallelized for specific hardware? How can batch jobs be structured to avoid overwhelming resources? Which simulators are best suited for validating solutions? And, when transitioning from simulation to real hardware, how much of the workflow needs modification?

In this work, we address these challenges by developing an automated, massively parallelized workflow for the Quantum Approximate Optimization Algorithm (QAOA)~\cite{qaoa2016suprem}. QAOA has become a leading candidate for solving combinatorial optimization problems on near-term quantum devices, thanks to its hybrid quantum-classical design and compatibility with noisy intermediate-scale quantum (NISQ) hardware. However, applying QAOA to large-scale problems often requires more qubits than currently available. One promising solution is problem partitioning, where a large problem is decomposed into smaller subproblems that can be solved independently.

Our workflow incorporates three essential components: (1) Problem decomposition, which partitions a constraint graph into manageable subgraphs; (2) Efficient subproblem generation; and (3) Automated batch job simulation, enabling the execution of diverse circuit sets at scale. We present a software library that automates these processes, reducing the user's workload and facilitating the efficient generation and execution of QAOA batch jobs through a unified interface.

On the simulation side, we develop a common interface that integrates multiple simulation engines. A key feature of this interface is its ability to act as a single point of entry for diverse simulators and simulation types. It automates the selection of the most suitable simulator for each circuit within a batch job, optimizing execution based on circuit characteristics such as complexity, size, and entanglement. Managing interoperability between different simulators is challenging due to variations in input formats, capabilities, and memory requirements. Our approach abstracts these differences, allowing users to run circuits on different backends without manual intervention, and maintaining performance of the simulator.

To ensure accessibility, we deploy these services through a cloud-based infrastructure. This system allows users to execute QAOA workflows without needing access to local high-performance computing resources. The cloud platform manages batch job submissions, allocates computational resources dynamically, and provides seamless scaling based on demand. This setup enables users to run large-scale QAOA experiments without worrying about the underlying infrastructure, ensuring efficient resource utilization and optimized performance.

The remainder of this paper is organized as follows: Section~\ref{sec:qaoa} details our automated parallelization workflow for QAOA and introduces Divi, our software library for managing this process. Section~\ref{sec:composer} discusses Maestro, our interface for integrating multiple simulators, and presents benchmark results. Section~\ref{sec:cloud} outlines the design of our cloud infrastructure, highlighting how it handles large-scale workloads. Finally, Section~\ref{sec:conclusion} concludes the paper and discusses future research directions.

\section{Parallelization Workflow for QAOA}\label{sec:qaoa}
To simplify the execution of large-scale QAOA problems on distributed systems, we develop a workflow and software package that automates the parallelization process. A key question that arises is whether problem partitioning affects solution quality. We first evaluate this on the MaxCut problem, detailing our approach to partitioning. Beyond partitioning, we introduce parallelization at the optimization step, integrating these components into our software library, Divi. Divi generalizes the workflow for arbitrary graph-based optimization problems, automates batch job generation, and provides efficient result aggregation for QAOA.

\subsection{Constraint-Graph Partitioning for QAOA}

An integration of graph partitioning techniques with a quantum resource-aware approach is essential. To address the challenges posed by large-scale optimization problems within the constraints of near-term quantum computing, a flexible methodology that balances both the problem's specific characteristics and the hardware's limitations is required. A balanced and adaptive strategy is crucial for optimizing performance and achieving practical advancements in the current quantum computing era.

In our methodology, we employ spectral partitioning to decompose a large graph into smaller subgraphs, enabling the application of QAOA to solve the MaxCut problem on quantum processors with practical limitations. This carries over to simulators who also have strict limitations in qubit counts. Spectral partitioning methods are advantageous for partitioning graphs as they balance subgraphs' sizes effectively while preserving the overall structure of the principal graph. In our implementation, rather than employing spectral bisection, we use k-means clustering to partition the graph using the Python library \textit{scikit-learn}, specifically its SpectralClustering algorithm~\cite{scikit-learn}. This enables us to partition based on the desired number of clusters, which is constrained by the qubit size that the subgraph must adhere to. This approach helps to mitigate the information loss associated with recursive bisection methods. 

An additional benefit of this method is its ability to group highly interconnected nodes into the same subgraph, regardless of node indices. This approach clusters densely connected nodes in close proximity, reducing the number of qubits that entanglement needs to propagate across---a significant advantage for qubit modalities with low connectivity. By reducing SWAP gates, this method addresses quantum processors' restricted operation due to decoherence and connectivity limitations, optimizing circuits for large graph partitioning on real hardware.

However, spectral graph partitioning methods have limited generalization as they do not consistently guarantee evenly sized subgraphs across all types of principal graphs. Specifically, for randomly generated graphs, this approach often results in an uneven distribution of nodes, where one subgraph may contain a disproportionately large number of nodes while others comprise significantly fewer. To address this limitation, we explored a novel methodology that integrates community detection clustering as a preprocessing step. This approach involves analyzing community structures within the graph to determine the size of communities associated with each node. The community size information is then embedded into the spectral clustering process to encourage a more balanced distribution of nodes, particularly ensuring that nodes from larger communities are dispersed across subgraphs. Subsequently, the k-means clustering algorithm is applied to identify partitions that are sensitive to community size variations. While this method yielded only marginal improvements in partitioning outcomes, it highlighted promising directions for future research into more sophisticated graph partitioning techniques.

\begin{figure}[ht]
    \centering
    \includegraphics[]{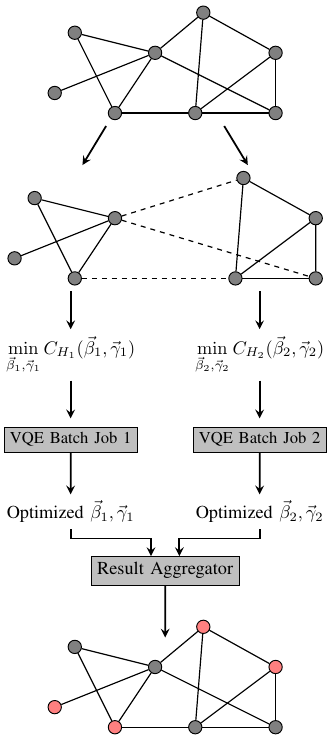}
    \caption{The parallelization flow used for QAOA with VQE. The flow begins with a principle graph which gets partitioned. From there, individual optimization programs are executed using VQE for finding the ground state of $H_i$. The optimal parameters are used to recover the optimal solution of each subproblem, which aggregate to form the final solution. }
    \label{fig:qaoa-flow}
\end{figure}

Once the graph is partitioned, each graph is optimized independently. The results of each sub-problem are then aggregated, resulting in one final result. End-to-end, the process for generating the set of partitioned graphs is depicted in Fig.~\ref{fig:qaoa-flow}. The figure illustrates a parallelization strategy for the QAOA using Divi's Variational Quantum Eigensolver (VQE) for optimization. The approach leverages graph partitioning, as explained, to break down a complex optimization problem into smaller, independent subproblems, which are solved separately before recombining the results.

\subsection{Partitioned Optimization Evaluation}

To evaluate the performance of partitioned QAOA, we use the MaxCut problem on various graph structures of varying size. The MaxCut problem entails partitioning a graph into two sets to maximize the number of edges between them. The MaxCut problem has applications in logistics, network design, and machine learning~\cite{maxcut2018app}. Additionally, the MaxCut problem serves as a standard benchmark for evaluating optimization algorithms due to its NP-hard complexity and will be used as the primary performance metric in our results.

When solving MaxCut with QAOA, each graph node corresponds to a qubit, totaling $n$ qubits for $n$ nodes. Edges are represented by controlled-$Z$ ($CZ$) gates, entangling the qubits. This setup encodes the problem into a quantum circuit, with each qubit's state ($|0\rangle$ or $|1\rangle$) representing the node's set assignment.

The QAOA circuit alternates between two primary operations: the cost Hamiltonian ($H_C$) and a mixer Hamiltonian ($H_M$):

\begin{align}
    H_C = \sum_{(i,j) \in E} \frac{I - Z_i Z_j}{2} && H_M = \sum_{i=1}^n X_i   
\end{align}

where $Z_i$ and $Z_j$ are Pauli-$Z$ operators acting on qubits $i$ and $j$. The cost Hamiltonian $H_C$ is implemented using $CZ$ gates for each edge in the graph and the mixer Hamiltonian $H_M$ using Pauli-$X$ operator on qubit $i$. The mixer enables transitions between states, allowing the algorithm to explore the solution space effectively.

Each layer of the QAOA circuit includes two operations: evolving under the problem Hamiltonian with parameter $\gamma$ and under the mixer Hamiltonian with parameter $\beta$. For $p$ layers, the QAOA circuit is:
\begin{equation}
U(\vec{\beta}, \vec{\gamma}) = \prod_{l=1}^p U_M(\beta_l) U_C(\gamma_l)  = e^{-i \beta_l H_M} e^{-i \gamma_l H_C},  
\end{equation}
where $\vec{\beta} = (\beta_1, ..., \beta_p)$ and $\vec{\gamma} = (\gamma_1, ..., \gamma_p)$ represent the variational parameters for each layer, which are iteratively optimized to maximize the expected value of the cut.

The optimization process involves preparing the quantum state $|\psi(\vec{\beta}, \vec{\gamma})\rangle$ and measuring the qubits to compute the cost. The parameters are adjusted using classical optimizer to minimize the expectation value of $H_C$, 
\begin{equation}
    \expval{H_C}{\psi(\vec{\beta}, \vec{\gamma})},
\end{equation} 
effectively maximizing the cut value. With this approach QAOA enables an efficient exploration of the solution space to approximate the optimal solution to the MaxCut problem.

\begin{figure}[ht]
    \centering    
    \includegraphics[]{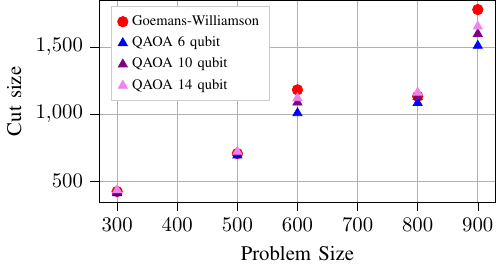}
    \caption{Partitioned QAOA evaluated in comparison to the Goemans-Williamson solver for varying sized problems. }
    \label{fig:scale-optimality}
\end{figure}

To assess the performance of partitioned QAOA against classical optimization methods, we conduct a comparative analysis using the MaxCut problem, where the cut-size serves as the primary metric for solution quality, where the higher the cut size, the better the result. Fig.~\ref{fig:scale-optimality} illustrates the comparative outcomes between a hybrid approach (graph partitioning followed by QAOA execution under qubit-count constraints) and the standalone Goemans-Williamson (GW) algorithm on circulant graphs. GW is an approximate solver that provides a guaranteed performance ratio of at least 0.878 for the MaxCut problem through semidefinite programming relaxation~\cite{gw1995}. We use this solver as our classical benchmark due to its computational efficiency and well-established theoretical guarantees, making it a standard reference point for evaluating quantum approaches to combinatorial optimization problems. 

The data demonstrates QAOA achieves parity with GW for small-scale graphs of circulant type ($n \lessapprox 500$), but diverges progressively as problem size increases, revealing an optimality gap. Crucially, QAOA implementations with larger subproblem dimensions (e.g., with qubit counts $n=14$ vs. $n=6$) exhibit enhanced approximation ratios, narrowing the GW performance gap. This empirical correlation underscores an inverse relationship between partition-induced subproblem reduction and solution fidelity---a trade-off inherent to divide-and-conquer solution strategies.

\begin{figure}[ht]
    \centering
    \includegraphics[]{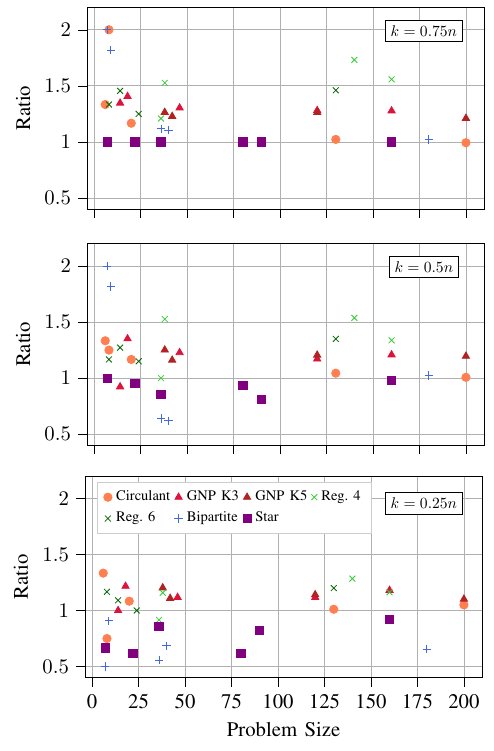}
    \caption{GW algorithm performance comparing the output (i.e., the cut-size) of the MaxCut problem using graph partitioning and unpartitioned graph. We show the ratio between the partitioned solution and the unpartitioned solution. Each of the plots shows the performance with different partition sizes $k$, where a larger $k$ represents more partitions used.}
    \label{fig:part-dist}
\end{figure}

Fig.~\ref{fig:part-dist} further quantifies this partitioning penalty through controlled experiments applying the GW algorithm to partitioned graph instances.  Distinct markers represent seven graph classes provided by HamLib~\cite{Sawaya_2024}: $\text{circulant}$ graphs, $\text{Erdős-Rényi}$ graphs with 3-clique constraints (GNP K3), 5-clique constraints (GNP K5), 4-regular graphs (Reg. 4), 6-regular graphs (Reg. 6), bipartite graphs, and star graphs. We systematically evaluate partitioning effects across three relative scales: $k = 0.25n$, $k = 0.5n$, and $k = 0.75n$, where $k$ denotes subproblem size and $n$ the total node count. 

The metric of interest, plotted on the $y$-axis, quantifies the normalized cut size $\tilde{C} = C_{\text{partitioned}} / C_{\text{unpartitioned}}$ where $C_{\text{partitioned}}$ represents the GW solution for partitioned subproblems and $C_{\text{unpartitioned}}$ the direct GW solution, which we call the cut-size ratio. Separate subplots analyze each partitioning scale, revealing that the magnitude of solution quality loss shows strong correlation with partition size $k$. For the $k=0.75n$ (the top plot), $\tilde{C}$ is mostly greater than $1$ indicating good relative solution quality for partitioned problem instances. But for $k=0.5n$ (the middle plot) $\tilde{C}$ begins to fall below $1$. And with $k=0.25n$ (the third plot), the majority of $\tilde{C}$ values are less than $1$, indicating superior solution quality for unpartitioned instances.

This empirical validation establishes subproblem size preservation as a critical control parameter for balancing quantum resource constraints against computational optimality in hybrid quantum-classical architectures, explicitly demonstrating the trade-off between quantum resource requirements and computational optimality (measured through approximation ratio preservation) consistently across diverse graph topologies.

Overall, the results show that in these graph structures, especially as the problem size increases, partitioning does not diminish the optimization quality in comparison to the GW algorithm. This allows larger problem instances to run, such that if QAOA is used, one can expect the results to maintain a good overall solution quality.  

\subsection{Divi: Automated Parallelization of Quantum Algorithms}\label{sec:divi}

As quantum computing hardware advances, the complexity of quantum algorithms and their execution requirements increase accordingly. Many quantum algorithms, particularly hybrid quantum-classical algorithms such as QAOA, VQE, or quantum machine learning, are well structured to be parallelized and distributed across hybrid computing infrastructure. To leverage this, it requires an understanding of the underlying structure of the algorithms and a software infrastructure to perform the various tasks involved to generate and aggregate parallelized quantum programs.

To address these challenges, we have developed Divi, a Python library designed to streamline the execution of large-scale quantum algorithms by automating the parallelization and batching of quantum programs. Divi is designed to integrate seamlessly with cloud infrastructure, automating away the complexity of parallelizing, batching, and aggregating quantum programs. Divi brings the following key features:

\textbf{(1) Automatic Parallelization of Quantum Programs.} Divi simplifies the generation and parallelization of quantum programs by automating the decomposition of large computational workloads into smaller, independently executable tasks. This ensures that problem complexity is reduced while maintaining solution quality, using the best methods to do so, offloading these task from the user. It is designed to do this for a large set of hybrid quantum programs, and continues to expand.

Divi follows a hierarchical design, where a principal problem is decomposed into unit subproblems, each requiring further processing. Currently, our focus is on optimization problems, which are broken down to enable parallelized optimization using Monte Carlo methods or parallelized genetic algorithms---an area we continue to explore. Additionally, serial optimizers such as the Nelder-Mead method are incorporated to reduce the number of circuits required to perform optimization, although requiring more job creations. 

Some algorithms perform better or worse depending on various decisions, and so Divi allows users to add decisions all at once to generate additional jobs. Despite parallelized optimization, iterative updates remain necessary. Divi automates the advancement of iteration steps for supported optimizers, handling expectation value computations and accommodating multiple measurement bases. This reduces the complexity for users while ensuring accurate optimization. It also handles fallback cases when failures occur, allowing the user to start from the last successful computation instead of starting from the beginning as most standard libraries are currently implemented.

Once subproblems are executed and optimization is complete, Divi aggregates results to construct the final solution. Its hierarchical structure allows parent programs to systematically collect and integrate subproblem results, ultimately providing a single, consolidated output to the user. It automates job tracking with the cloud interface, ensuring all results completed before aggregation starts.

\textbf{(2) Efficient Batch Job Generation.} Once subprograms are created, their respective circuits must be generated in a format that can be serialized for execution. To achieve this, we use routines from Pennylane~\cite{bergholm2018pennylane} to generate circuit abstractions. Pennylane provides a flexible and efficient representation of quantum circuits while allowing for low-level control over circuit instructions and measurements. Additionally, it includes functionality for converting circuits to QASM.

However, we found the QASM conversion in the latest Pennylane release inefficient for batch QAOA tasks. When computing the expectation value of a Hamiltonian---vital in VQE and QAOA---measurements must be in the basis of each Hamiltonian term. Ideally, the QASM conversion should generate one program per term, but the existing routine produces only a single QASM program, leading to incomplete execution. To fix this, we implemented a corrected conversion routine in Divi, ensuring the correct number of QASM programs are generated.

When converting abstract circuits to QASM, high-level circuits must be translated into QASM-compatible gate instructions. In QAOA, the Trotterization step approximates the time evolution of the Hamiltonian. Naively generating batch circuits causes redundant Trotterization processes for each circuit, greatly increasing runtime. Divi addresses this by using symbolic parameter representations, performing Trotterization once per program to create a low-level \enquote{scaffold} QASM. Parameters are then substituted (\enquote{grounded}) with specific values.

\begin{figure}[ht]
    \centering    
    \includegraphics[]{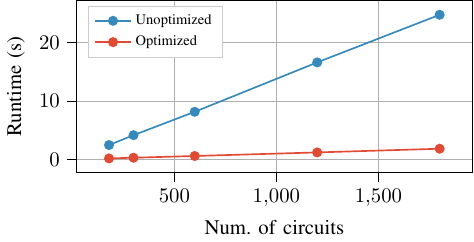}
    \caption{Time it takes to generate batch circuits for QAOA for 5 nodes and varying the number of samples with MC optimization.}
    \label{fig:batch-circuit-gen}
\end{figure}

This approach leads to significant speedups in performance, as shown in Fig.~\ref{fig:batch-circuit-gen}. Here we see the naive approach to generate the collection of QASM circuits for a problem of only 5 nodes can take more than 20 seconds to generate 1,500 circuits on a personal computer, where with Divi's optimization, the time to generate the circuits is reduced by a factor of 13. This problem is not isolated to Pennylane, and any circuit programming language that goes from an abstraction to QASM using Trotterization will face the same issue if these optimizations are not considered.

\textbf{(3) Cloud Interfacing.} Divi is currently designed to run at the user side and is designed to seamlessly interface with our cloud services. By integrating with our cloud API, Divi automates the submission, tracking, and retrieval of results for batch quantum jobs, ensuring successful execution and fallbacks for the cases of error.

Once a quantum workload is partitioned and prepared, Divi submits batch jobs to our cloud infrastructure via the API service. It tracks the execution progress of subprograms, continuously monitoring job states and retrieving partial results as they become available. If a job fails or a connection is lost, Divi is designed to apply fallback mechanisms, ensuring that previously completed computations are saved and resumed without requiring a full restart. This approach prevents unnecessary recomputation. From there, the user gains a set of additional benefits, as will be described in Section~\ref{sec:cloud}.

\section{Simulating Batches of Circuits}\label{sec:composer}
Simulating quantum circuits is crucial for developing and validating quantum algorithms, particularly in this era of quantum computing. As quantum hardware remains limited, classical simulation provides a vital testbed, but its computational cost scales exponentially with qubit count and entanglement. To address this, various simulation techniques have been developed, each balancing accuracy, memory usage, and efficiency. This section reviews the strengths and limitations of current approaches, including state-vector, matrix product state (MPS), and tensor network simulators. We then introduce our approach, Maestro, designed to optimize performance through automation and parallelization, and benchmark for performance.

\subsection{Quantum Circuit Simulation Methods}

Quantum circuit simulation uses various methods, each offering trade-offs between accuracy, scalability, and computational efficiency. State-vector simulation is the most direct approach, representing the quantum state as a complex vector of dimension $2^n$, where $n$ is the number of qubits. This method allows exact simulation and full access to state amplitudes, making it ideal for debugging and analyzing quantum algorithms. However, its exponential memory requirements limit scalability---simulating more than 30 qubits can require hundreds of gigabytes of memory---making it impractical for large systems but effective for small- to medium-scale circuits.

Matrix Product State (MPS) simulation offers a more scalable alternative by representing quantum states as chains of tensors with bounded bond dimensions~\cite{vidal2003efficient, schollwock2011density, orus2014practical}. This approach efficiently handles circuits with limited entanglement, significantly reducing memory requirements compared to state-vector methods. However, its efficiency depends on the circuit’s structure; highly entangled circuits lead to increased bond dimensions, limiting scalability. MPS simulators are particularly suited for one-dimensional systems, shallow circuits, and variational algorithms with constrained entanglement.

Tensor network simulators generalize MPS by using interconnected tensors to represent quantum states, accommodating more complex circuit structures~\cite{orus2014practical, schollwock2011density}. These methods can efficiently simulate circuits with bounded entanglement and are valuable in areas like quantum error correction and many-body physics. However, as entanglement increases, tensor contractions become computationally expensive, limiting scalability. Despite this, tensor networks offer a balance between accuracy and efficiency for structured circuits.

To accelerate simulations, some frameworks leverage Graphical Processing Units (GPUs)\cite{qiskit2024, bayraktar2023cuquantum}. GPU-based simulators can perform parallelized matrix operations and tensor contractions, significantly speeding up large-scale simulations. However, they introduce challenges like memory bottlenecks and data transfer overheads, which can negate performance gains, especially in smaller circuits or when frequent communication is needed\cite{faj2023quantum}. While GPUs excel in specific scenarios, they often require careful memory management to outperform CPU-based simulations, in particular for smaller scale circuits.

Choosing the right simulation method involves balancing precision, scalability, and computational resources. State-vector simulators are ideal for exact results in small circuits, while MPS and tensor networks extend capabilities to larger or structured systems. GPU-accelerated methods can enhance performance for large, parallelizable workloads but may struggle with smaller or memory-intensive tasks. Additionally, factors like hardware configuration---cores, memory, and architecture---impact performance, complicating simulator selection, especially in batch processing scenarios with diverse circuits.

Efficiently handling multiple simulation shots is also crucial. Repeating circuit executions ensures statistical accuracy, but poor handling can introduce computational overhead. Optimizing for multi-shot performance is essential for tasks like quantum program validation, where balancing fidelity and efficiency is critical. This adds another layer of complexity to simulator selection, particularly when considering large-scale batch jobs.

The simulation ecosystem includes diverse tools with varying capabilities. IBM’s Qiskit Aer stands out for its flexibility, supporting state-vector, MPS, and GPU-accelerated simulations~\cite{qiskit2024}. It can automatically select suitable methods based on circuit structure and system memory, optimizing for specific tasks like stabilizer circuits. However, this automated selection is limited to Qiskit’s ecosystem and does not  optimize for execution performance across different hardware parameters or other simulators outside of Qiskit.

Given the wide range of simulation methods, providers, and hardware configurations, selecting the optimal simulator for a given task is complex~\cite{jamadagni2024benchmarking}. Different circuits benefit from different approaches, and no single method offers universal efficiency. To address this, a flexible simulation framework that interfaces with multiple backends is highly beneficial for adapting to diverse quantum algorithms and hardware. It can reduce runtimes according to the user's needs, accelerating development and potentially reducing energy consumption for simulating practical-scale quantum programs in HPC centers.

Maestro, our platform, addresses these challenges by integrating multiple simulation paradigms into a unified interface. It allows users to leverage the most appropriate simulator for their specific needs while maintaining interoperability across quantum hardware and software ecosystems. In the following section, we detail Maestro’s architecture and capabilities.

\subsection{Maestro: The Interface to Quantum Circuit Simulation}

As quantum simulation evolves, integrating diverse methodologies into a unified and efficient framework remains a significant challenge. Maestro addresses this by serving as a common interface that connects multiple quantum circuit simulators, preserving and, in some cases, enhancing their capabilities. Implemented as a C\texttt{++} library, Maestro aims to unify the quantum simulation ecosystem through a standardized interface, automate simulator selection, and optimize critical tasks like multi-shot execution and simulated distributed quantum computing.

One of Maestro’s core features is its ability to interface with various simulators through its intermediate representation input (e.g. OpenQASM). Each quantum simulator---whether it is a state-vector, MPS, tensor network, or hardware-specific simulator---can have its own representation and simulation paradigm, requiring conversion. Maestro introduces a translation layer that maps intermediate representations of circuits to its own representation, to then interface to other simulators. This enables users to run circuits across different simulation platforms without needing to adapt the circuit representation for each simulator they want to use manually. Maestro also performs circuit analysis and compression to optimize efficiency. Currently, Maestro supports QCSim\cite{QCSim} and Qiskit Aer\cite{qiskit2024}, but the framework is designed to be easily extensible, using class interfaces. Adding support for new simulators involves implementing the interface, simplifying integration.

Maestro provides an automated simulator selection mechanism to optimize performance. When a circuit is to be run, at runtime, Maestro has two ways to decide which simulator to use. Firstly it can execute a single shot across multiple supported simulators in parallel, records execution times, and selects the right backend for the remaining shots, with an understanding of the multi-shot optimization performance for each. This method is simple and flexible in that it works regardless of software updates to the simulators and the underlying hardware. The downside is, running one shot on many simulators occupies computing resources and can delay the computation in some case.

The second method we use, as an alternative, uses simulation algorithm complexity with circuit information in addition to hardware features (determined at compile time) to make the decision. At compile time, the constants for the algorithm complexity are determined based on the hardware, and then used to select the right simulator for the input circuit. Regression methods are used to extrapolate runtime. This method is very efficient, as it is essentially a lookup task, but it can be challenging to determine the complexity of the simulation algorithm for each simulator, especially when they use parallelization and multi-threading, and requires care. This is an ongoing research direction which we aim to explore more deeply.

Another core feature of Maestro is its optimization of multi-shot execution, a process where quantum circuits are run thousands of times to build statistically meaningful results. Many existing simulators either lack efficient multi-shot execution or do not implement it at all, leading to significant runtime overhead. Maestro addresses this by optimizing how circuit executions are managed. It stores essential information---such as the observable being measured and the final quantum state---allowing the most computationally intensive steps to run only once. This approach significantly reduces redundant computations. Maestro also handles complex cases, such as circuits with mid-circuit measurements, essential for testing error correction, by retaining state information up to the measurement point. This prevents unnecessary recalculations of earlier circuit sections, further improving execution efficiency.

For simulating distributed quantum computing, Maestro uses a \enquote{composite} simulation approach tailored to the nature of distributed circuits. In distributed quantum circuit, circuits are partitioned across multiple nodes, leading to fluctuating system sizes as qubits become entangled and disentangled between devices during execution. Maestro addresses this by dynamically adjusting the simulator’s Hilbert space: expanding it when entanglement occurs and contracting it after measurements. This adaptive approach minimizes memory usage and computational overhead, enhancing performance and enabling efficient parallelization. Given that entanglement and disentanglement occur frequently in distributed quantum algorithms, this method can offer significant speedups.

Maestro represents a substantial advancement in creating a more flexible and efficient quantum simulation ecosystem. By unifying diverse simulation methodologies, automating backend selection, and introducing optimizations for multi-shot execution and distributed quantum computing, Maestro simplifies the simulation process for end-users who are often interested in the best result as fast as possible, without having to understand the ins and outs of each simulation platform and methodology. 

\subsection{Benchmarking Results}

\begin{figure*}
    \centering    
    \begin{subfigure}{0.48\textwidth}
        \centering                
        \includegraphics[]{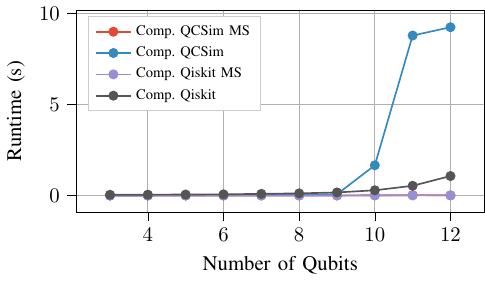}
        \caption{State vector simulation for 5,000 shots with and without multi-shot optimization.}
        \label{fig:state-vec-1shot}
    \end{subfigure}
    \hfill
    \begin{subfigure}{0.48\textwidth}
        \centering                
        \includegraphics[]{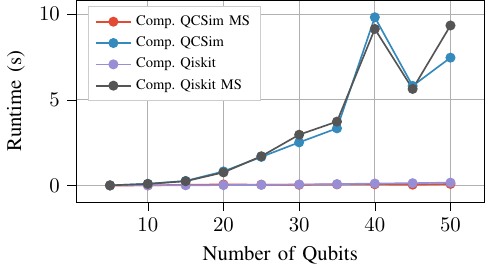}
        \caption{MPS simulation for 100 shots with and multi-shot optimization.}
        \label{fig:state-vec-10kshot}
    \end{subfigure}
   \caption{Comparison of state vector simulation with and without Maestro for different shot counts.}
\label{fig:bench-shots}
\end{figure*}

We focus on how integration with Maestro influences performance and usability of a simulator. So far, Maestro interfaces with QCSim~\cite{QCSim} and Qiskit Aer~\cite{qiskit2024}, and we examine the impact on simulation efficiency. Many benchmarking results focus on timing aspects for single shot performance or on a particular circuit structure. Understanding the factors that influence simulator performance is for real-world applications is crucial, and it is practical to understand how simulators handle batch jobs with varying numbers of shots, circuit sizes, depths, and other features. Effectively handling dynamic input scenarios can yield significant performance gains, highlighting the importance of adaptive simulation strategies. 

\begin{figure}[ht]
    \centering
    \begin{subfigure}{0.48\textwidth}
        \centering        
        \includegraphics[]{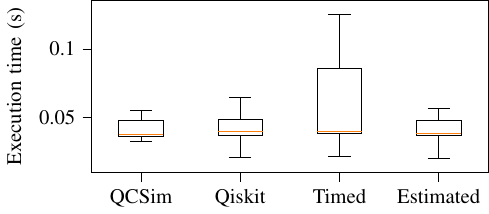}
        \caption{Execution times for a batch job of circuits.}
        \label{fig:batch-job}
    \end{subfigure}
    
    \vspace{1em}
    
    \begin{subfigure}{0.48\textwidth}
        \centering
        \includegraphics[]{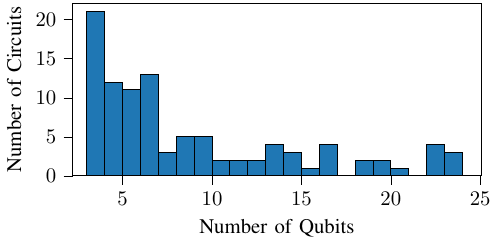}
        \caption{Distribution of circuits executed.}
        \label{fig:circ_dist}
    \end{subfigure}
    \caption{Execution time distribution for a batch job of circuits. All circuits run for 5,000 shots.}
    \label{fig:bench-batch}
\end{figure}

We benchmark different tasks for Maestro wrapping QCSim and Qiskit. 1) The runtime  without multi-shot optimization; 2) The runtime with multi-shot optimization; and 3) The performance of batch jobs with automatic simulator selection. We use benchmarking circuits from the MQT Bench archive~\cite{quetschlich2023mqtbench}, specifically the QAOA circuits and for circuits with more qubits, since the QAOA circuits are limited to 16 qubits, the quantum neural network circuits, as they tend to have similar structure to QAOA circuits. The infrastructure we use to execute the benchmarks is a 16-core CPU with 128 GB of RAM. Figures~\ref{fig:bench-shots} and \ref{fig:bench-batch} show the results.

In Fig.~\ref{fig:state-vec-1shot} and \ref{fig:state-vec-10kshot} we evaluate the difference between using multi-shot optimizations vs. not using them. We can see in Fig.~\ref{fig:state-vec-1shot}, already with 12 qubits, using the optimized execution, the runtimes grow significantly. Most drastically, Maestro's wrapping of QCSim without any multi-shot optimization takes nearly 10 seconds to execute 5,000 shots. Including these optimizations, the runtime is brought down to 0.007 seconds. These optimizations are applied in an abstract way, and so any simulator we add in the future missing these enhancements would see these benefits. In Fig.~\ref{fig:state-vec-10kshot}, a similar behavior is observed for MPS simulation. Optimization here is handled similarly, and the effect is the same.

In Fig.~\ref{fig:bench-batch}, we analyze the execution time distribution for quantum circuit batch jobs across various simulation methods. We compare QCSim statevector, Qiskit statevector, timed estimation, and an estimation routine using randomly sampled circuits of up to 23 qubits from quantum neural network benchmarks, with 5,000 shots per circuit execution. Our results indicate that the simulator methods perform similarly on average, though with notable differences in distribution. The estimation method and Qiskit demonstrates the fastest execution time overall the circuits, while QCSim and the estimated version exhibits the smallest maximum execution time. Timed simulation, despite comparable average performance, shows a wider spread primarily due to resource-intensive initialization. These techniques warrant further investigation in future work.
 
Overall, automatic selection of simulators is a non-trival task, but evidence shows that having this feature can result in a better overall performance. When there are batch jobs consisting of circuits with a high number of qubits, it can cause overuse of computing time and resource starvation for other users if the incorrect choices are made. This is a problem we will focus on in depth in follow up research.

\section{Infrastructure for Batch Quantum Programs}\label{sec:cloud}
The goal of this work is to develop software tools that simplify the validation and execution of practical-scale quantum programs. We have outlined how job batching is enhanced, allowing users to specify optimization requirements for QAOA while Divi automatically partitions the problem and handles execution of jobs. These jobs are simulated through Maestro, which optimizes multi-shot simulations, selects the best-performing simulators, and abstracts the complexity of using many simulator from the user. The final component of our approach is providing robust cloud infrastructure, allowing users to access high-performance computing resources and QPUs without the need for direct ownership---an especially important consideration as quantum datacenters continue to emerge.

In this section, we present our cloud-based infrastructure, which serves as a critical middleware layer between applications and hardware. By integrating widely used cloud services, our system manages the various processing steps required for seamless execution. We describe the overall design of our infrastructure and show an example of how it is used, demonstrating the vast reduction in end-user code, the simulation time reduction, and simplicity added.

\subsection{Implementation}

To deploy the infrastructure, we deploy our services as elastic container services on Amazon Web Services (AWS), leveraging its scalability, monitoring, and orchestration capabilities. This ensures dynamic resource allocation, optimizing computational efficiency while minimizing overhead. AWS provides real-time monitoring and fault tolerance, improving system resilience and reliability. Additionally, geographic distribution across multiple regions reduces latency, particularly for cloud controllers, and enhances accessibility to both quantum and classical computing resources. The flexibility of containerized services also allows components of the system to be deployed on-premise or across hybrid cloud environments, ensuring adaptability to various computing infrastructures. Overall, the system is designed to be scalable, resilient, and deployment-agnostic, supporting both cloud-based and localized execution. 

\begin{figure}[ht]
    \centering    
    \includegraphics[]{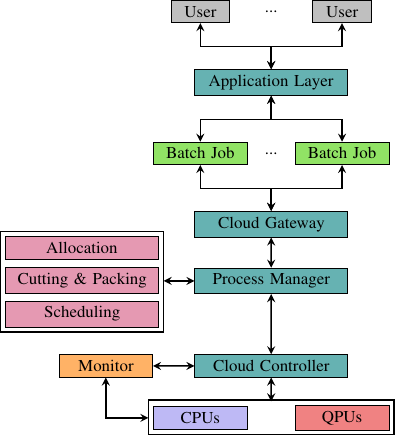}
    \caption{The infrastructure of services used in the workflow.}
    \label{fig:services}
\end{figure}

Fig.~\ref{fig:services} depicts the general end-to-end infrastructure. The end-to-end implementation contains all necessary components for automated, distributed simulation of quantum programs. The cloud gateway serves as the entry point, managing a queuing system and relational database that tracks circuit information, execution schedules, user data, and real-time computing infrastructure updates. Jobs originate from the application layer, implemented through Divi, which handles parallelization, batch job creation, and interacts with the cloud gateway via an API to monitor job execution.

The process manager oversees the job queue and triggers execution workflows using a Simple Queue Service (SQS). Upon receiving a quantum program, it selects available computing nodes from the infrastructure database and initiates a scheduling event. The scheduling process is managed by a separate service, which determines execution rounds based on both circuit requirements and resource availability. Once the schedule is generated and stored in the database, the process manager forwards it to the cloud controller for execution. It also contains services for network-aware resource allocation and circuit cutting and packing---a topic we expand on in future work.

The cloud controller, implemented as a separate service, organizes circuits as compressed QASM data along with metadata and destination addresses for computing nodes. It communicates with the host nodes via HTTPS, interfacing with their API to transmit circuits for execution. Host nodes expose a status endpoint that updates the database with execution progress. The cloud controller monitors these interactions, ensuring job execution and retrieving measurement results.

For testing, we deploy host nodes as emulations of QPUs, deployed again as elastic services in AWS, mimicking the interaction of interfacing with a QPU, but running the circuits using Maestro. Because the circuits are given in QASM format, these can be straightforwardly swapped to real QPUs. We are able to spin up many instances of the host nodes, which we put behind a load balancer. In this way, we can easily scale the computing power of the system for simulation purposes. As long as a node exposes the few endpoints required, the computing device itself can be flexible. Further, by using Maestro in the emulators, it is straightforward to package up, with a single simulation library to install, and leveraging the suite of circuit simulators.

To demonstrate the flexibility of the infrastructure, we also deploy a service to connect to the AWS Braket platform~\cite{braket}. With this, we include the service as a possible endpoint to receive and execute quantum circuits, also acting as an emulated QPU, returning the statistics in the format needed for aggregation of the results. We use the simulation platform of Braket for this, but one could change the device to a real QPU if desired.

Upon execution, the emulated QPUs return measurement results and metadata (e.g., runtime, execution environment, etc.) to the process manager, which stores the data in the database. Since these results may correspond to cut or packed circuits, additional processing is required to reconstruct the original job outputs. Aggregation and marginalization steps are applied to recover full results, ensuring consistency with the original problem formulation. Once all circuits are processed, the job state updates from running to complete, triggering Divi to retrieve the measurement results for post-processing and application-level aggregation.

Overall, this architecture implementation enables a fully automated workflow for executing and simulating quantum programs over a distributed computing infrastructure. By leveraging cloud-based orchestration, parallelized job execution, and flexible deployment strategies for simulation, the system abstracts the complexity of resource management, scheduling, and data handling, allowing users to focus on algorithm development rather than execution logistics. The use of elastic services and hardware-agnostic interfaces ensures scalability and adaptability across both cloud-based and on-premise environments. This framework paves the way for efficient large-scale quantum computing experiments, providing a robust foundation for hybrid quantum-classical execution in real-world applications.

In the end, the amount of code left to the user amounts to the following Python code block using Divi:
\begin{tcolorbox}[colframe=gray!10, colback=gray!10, sharp corners, left=5mm, right=2mm]
\begin{lstlisting}[style=mystyle]
q_service = QoroService(API_KEY)
graph = user_graph
batch = GraphPartitioningQAOA(
    problem=QAOA.MAX_CUT,
    graph=graph,
    n_layers=2,
    n_clusters=5,
    optimizer=Optimizers.MONTE_CARLO,
    max_iterations=1,
    qoro_service=q_service,
)
batch.create_programs()
batch.run()
quantum_solution = batch.aggregate_results()
\end{lstlisting}
\end{tcolorbox}
With a circulant graph of 30 nodes, using increasing number of samples for the Monte-Carlo method, putting the end to end system together we get the following runtime results running Maestro over a network of five computing nodes. We use 5,000 shots for each circuit and compare to a local simulation using Qiskit Aer. The results are shown in Table~\ref{tab:runtime_results}. 

\begin{table}[h]
    \centering
    \begin{tabular}{|c|c|c|}
        \hline
        \textbf{Num. Circuits} &  \textbf{Local Sim. Time (s)} & \textbf{Cloud Sim. Time (s)}  \\ 
        \hline
        3,000  &  17.65  &  1.34   \\ 
        \hline
        6,000 &  22.16 & 2.55 \\ 
        \hline
        12,000  & 31.12 & 5.45    \\         
        \hline
    \end{tabular}
    \caption{Performance metrics for different problem sizes over the cloud service vs. locally.}
    \label{tab:runtime_results}
\end{table}

Our results demonstrate that the end-to-end platform significantly reduces simulation time through batch job parallelization and the estimator technique. The system demonstrates scalability regardless of size, with potential for further performance improvements through additional computing nodes and system enhancements.

\section{Outlook and Conclusion}\label{sec:conclusion}
In this work, we have established a comprehensive framework that fundamentally transforms the execution of large-scale QAOA workflows. Our approach addresses critical scalability barriers in quantum computing by automating complex decision-making processes that previously required significant user intervention. Through automated problem partitioning, efficient batch job execution, and a common interface for integrating multiple simulation engines, we have demonstrated how quantum workflows can be streamlined to maximize both scalability and usability.

Our integrated system consists of complementary components working seamlessly together. Divi automates the parallelization and batching of QAOA and VQE jobs through intelligent problem partitioning, with a clear path to address more problems. Maestro provides a unified interface to multiple simulation engines, dynamically selecting the optimal simulator based on circuit characteristics and hardware performance. Our Cloud Platform orchestrates distributed execution through a sophisticated scheduling system that efficiently allocates computing resources across heterogeneous clusters, including combinations of simulators and real devices.

This framework represents a critical step toward practically useful quantum computing by bridging the gap between theoretical algorithm design and real-world implementation and system engineering challenges. Building on this foundation, our immediate development priorities include implementing realistic noise simulation capabilities and integrating GPU-accelerated simulators to further enhance performance. 

The future of quantum computing requires the convergence of high-performance computing with quantum systems. In the near term, distributed quantum computing offers compelling advantages through multi-modal, multi-vendor computing networks, while quantum-connected quantum computers will eventually enable circuit scaling beyond what is possible with classical approaches alone. Realizing this vision demands robust software infrastructure that seamlessly coordinates diverse computing resources while abstracting complexity from users.

Through this work, we aim to catalyze standardization efforts and promote the potential for automatization in quantum computing that will accelerate the transition from theoretical quantum computing to practical quantum advantage in real-world applications.

\bibliographystyle{IEEEtran}


\end{document}